# High Gain Array Antenna With FSS for Vital Sign Monitoring Through the Wall

Rifa Atul Izza Asyari, Rezki El Arif, Wei-Chih Su, and Tzyy-Sheng Horng
*Department of Electrical Engineering, National Sun Yat-Sen University, Kaohsiung, Taiwan*

*Abstract*— In this research, a 2.4 GHz planar array antenna 2×2 combined with frequency selective surface (FSS) is proposed to obtain a high gain and high efficiency for improving the performance of vital sign monitoring through the wall using Self-injection-locked radar systems. The FSS is designed on a double-layer with unit cells of 9×7 for each layer. Modified Jerusalem cross-shaped of FSS is proposed to enhance the gain. The antenna array and additional FSS can obtain a gain of 15.2 dB and 82% of efficiency. the vital sign monitoring of the subject behind the wall has been successfully carried out.

*Keywords—SIL radar, high gain antenna array, FSS, and metasurface*

## I. INTRODUCTION

Recently, a real-time see-through wall technology has gained more attention due to its capability to detect, locate and monitor the people behind the wall [1]. Some papers have been proposed to implement the Doppler radar for such applications [1]-[4]. Since the 1970s, CW radar has been widely used for non-contact cardiopulmonary monitoring and this idea can be applied to search for victims under earthquake rubble. However, its major shortcoming is the need of isolating the transmitted (Tx) from the received (Rx) signals using either a circulator or separate Tx and Rx antennas. Hence, reducing the size and cost of integration of CW radar for medical applications encounters significant challenges. To deal with such a problem, the authors proposed a self-injection-locked (SIL) radar for motion detection and vital sign monitoring [5]. It gives advantages such as system simplicity, immunity, and sensitivity to stationary clutter caused by mismatched component and transmitter-receiver coupling in the environment.

High gain antennas are needed for advanced radar system as an essential component of the system [6]. Some efforts have been done to design a high gain antenna such as the horn antenna [7], and the planar antenna array [8]. Comparing to the horn antenna, the planar antenna array is relatively easier to fabricate and has a lower cost. However, designing a high gain antenna array needs a wider dimension of the antenna. Another technique to enhance the antenna gain is by implementing the frequency selective surface (FSS), allowing to maximize the radiation efficiency and enhancing the efficiency [8]. The combination of array antenna and FSS structure can be utilized as a high gain antenna to obtain a better performance of the radar system while maintaining a small dimension.

In this paper, a planar array antenna combined with FSS is implemented for SIL radar to monitor the vital sign of the subject through the wall. This combination provides a highly directional beam by avoiding power loss in other directions. Thus, it can provide a high gain antenna so the system has better sensitivity.

## II. ANTENNA THEORY AND DESIGN

Fig. 1 shows the planar antenna array that consists of the 2×2 radiator patch. In such arrays, the radiation pattern is the vector summation of the fields radiated from the individual antenna elements, resulting in the high gain radiation. Commonly, an antenna array desires the same excitation in each uniformly placed element which is separated with the same distance between the adjacent elements [9]. In this study, the 2x2 radiator patch is connected to the feedline by using copper vias with a diameter of 1mm.

As shown in Fig. 1(d). The radiation source of the antenna is reflected multiple times with reducing amplitudes between the metasurface and the ground. These multiple reflections increase the antenna gain consequently since the air gap "$H_2$" between the ground layer and the metasurface is corresponding to each other [10].

In pursuance of the characteristic of the incident wave of frequency selective surface, a proposed FSS based on the modified Jerusalem cross and phased array antenna is designed as shown in Fig. 1(b). The structure of FSS with a periodic array of 8×6 units are placed in front of the antenna as shown in Fig. 1(a). The FSS is designed using the FR-4 substrate which is inexpensive, having a thickness of 0.8 mm and a dielectric constant of 4.4. Table I describes the design parameter of the antenna and the FSS structure.

Moreover, after dealing with the proposed architecture of the antenna array and FSS then the antenna was fabricated on FR-4 substrate ($\varepsilon_r$ = 4.4, tan($\delta$) = 0.01) in multi-layer PCB. After that, the simulation and measurement results of the antenna design

Table I
DESIGN STRUCTURES OF THE ANTENNA ARRAY AND FSS

| $W_f$ | $L_f$ | $W_p$ | $L_p$ | $\varepsilon_r$ | via |
|---|---|---|---|---|---|
| 1.5 mm | 1.5 mm | 15 mm | 56 mm | 4.4 | 1 mm |
| $W_1$ | $W_2$ | $W_3$ | $W_4$ | $H_1$ | $H_2$ |
| 4.8 mm | 3.5 mm | 3.2 mm | 2 mm | 14 mm | 265 mm |

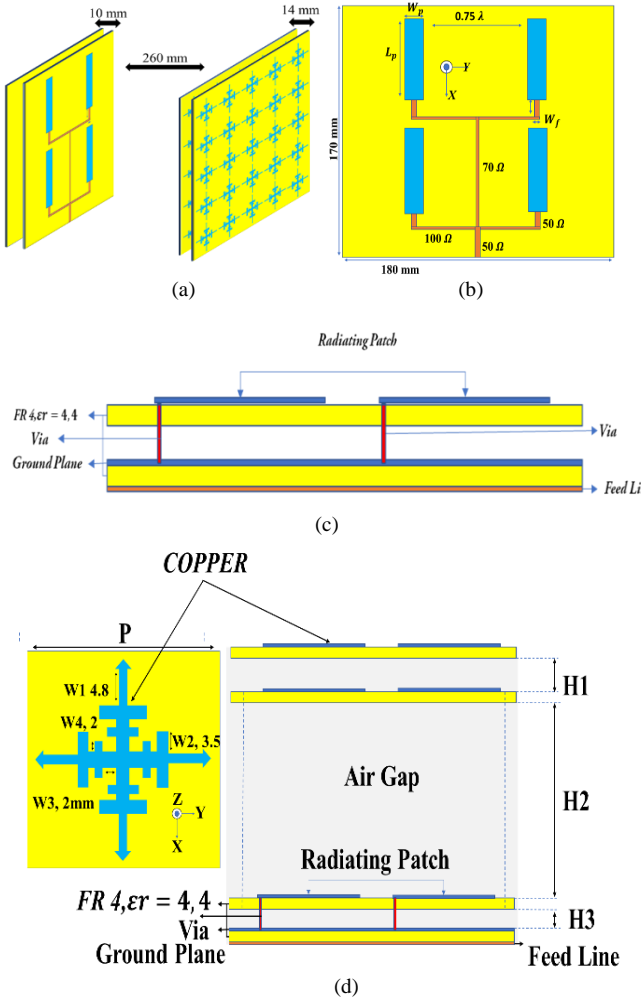

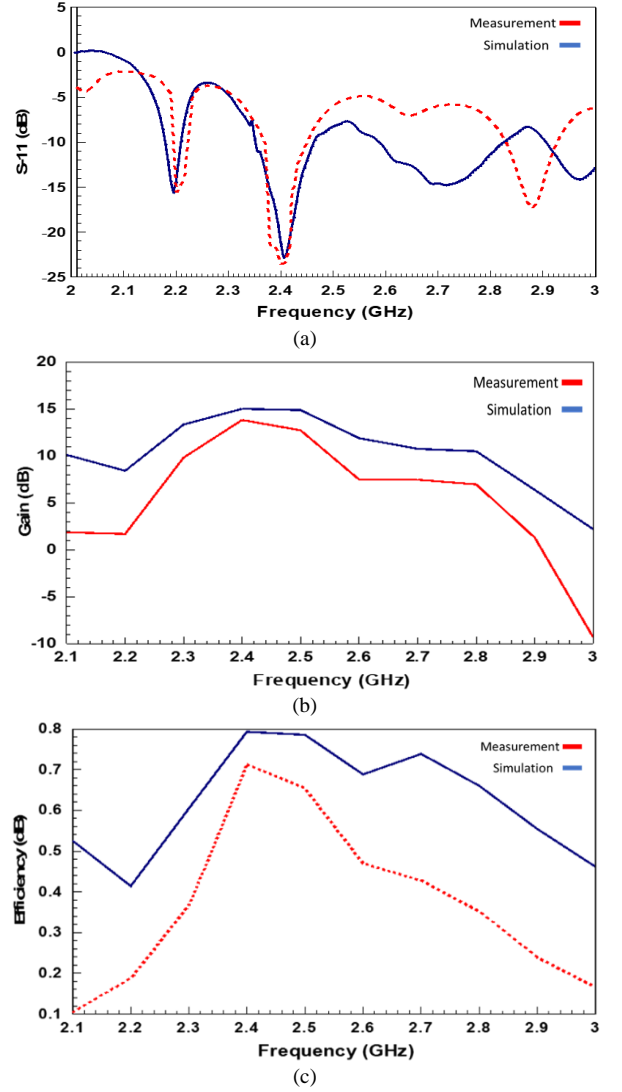

Fig. 1. The structure of the matasurface antenna. (a) 3D view of the array antenna with metasurface. (b) Array antenna with inset feed. (c) Array antenna double layer geometry. (d) Structure of the array antenna and FSS.

Table II
SIMULATION AND MEASUREMENT COMPARISON OF AN ARRAY ANTENNA WITH FSS

| Array 2x2 with FSS | S11 (dB) | Efficiency | Gain (dB) | Full beam width (degrees) | Half beam width (degrees) |
|---|---|---|---|---|---|
| Sim. | -24 | 82% | 15.2 | 8.43 | 4.21 |
| Meas. | -25 | 70% | 13.1 | 6.56 | 3.32 |

are compared. There are three fundamental parameters, first is the frequency of operation ($f_0$). The electrical dimensions of the unit cell is $0.14\ \lambda_0 \times 0.14\ \lambda_0$. Where $\lambda_0$ refers to the wavelength of the operating frequency of 2.4 GHz. The total dimension of the one-layer FSS is $200 \times 200$ mm$^2$. Fig. 2 shows the result of the simulations and measurements of the array antenna with the FSS structures. They show a good agreement between the simulation and the measurement results. The performance shows that the array antenna 2×2 with FSS is outperformed in terms of efficiency, bandwidth, and gain. The evaluated parameters are

Fig. 2. (a) Simulation and measurement of the antenna parameter (a) S-parameter. (b) gain. (c) Efficiency.

S11, gain, radiation pattern, efficiency. The return loss (S11) of a patch antenna array with 4 elements is presented in Fig. 2(a). The peak gain and the efficiencies are 15.1 dB and 82% at 2.4 GHz in Fig. 2(b) and (c), respectively. The radiation patterns of simulation and measurement are shown in Fig. 3. The simulation return loss is equal to -20.24 dB at the centre frequency of 2.4 GHz. More completed outcomes about the simulations and measurements data are shown in Table II.

## III. RADAR DESIGN

Fig. 4 shows the proposed diagram systems of the receiver using an IQ demodulator that can extract the modulated signal from the subject's vital signs. In this receiver, the proposed radar system consists of a phased array antenna with FSS and an injection-locked oscillator (ILO) based receiver. It has the antenna and FSS to radiate and receive the signal from the ILO. After the echo signal is received including the doppler information of the subject's vital sign, the signal is known as the injected signal that makes the ILO enter to the injection state. The echo signal is amplified by the low noise amplifier (LNA) with a gain of 19 dB. After the signal pass through the frequency demodulator, the baseband signal is obtained. The baseband signal was acquired by a DAQ. The software used to process the signal are the NI LabView software and the advanced design system (ADS) software.

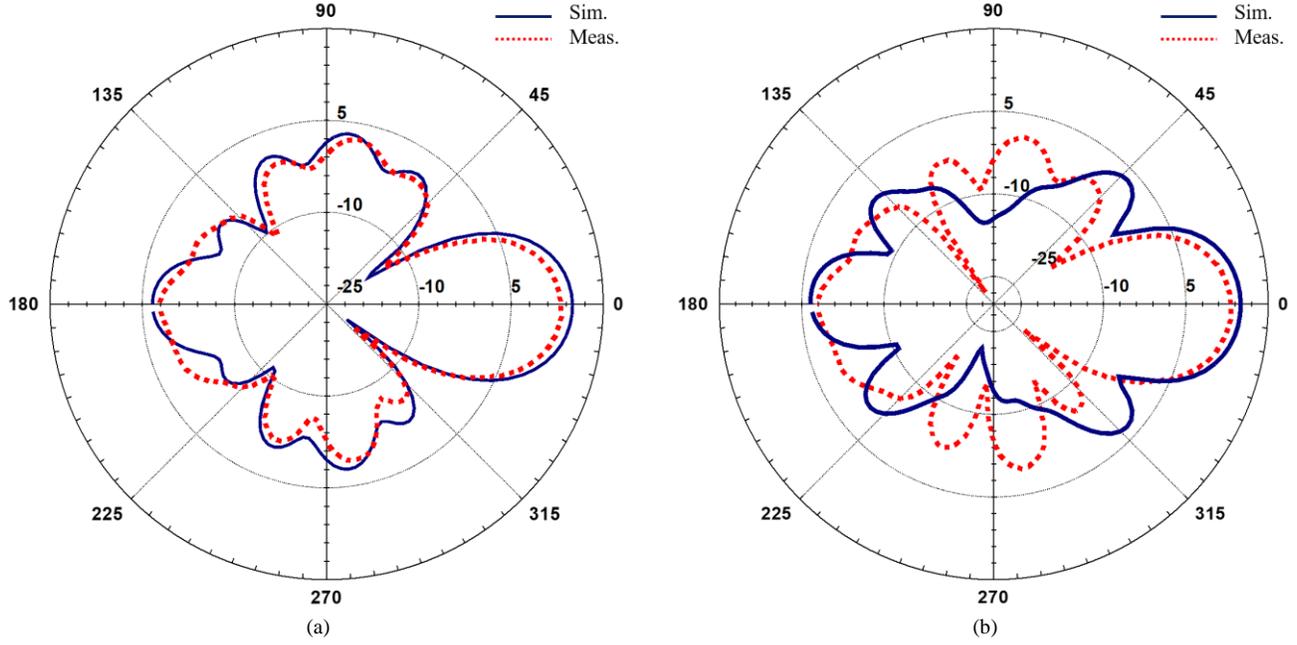

Fig. 3. Radiation pattern of the array antenna with FSS. (a) E-field. (b) H-Field.

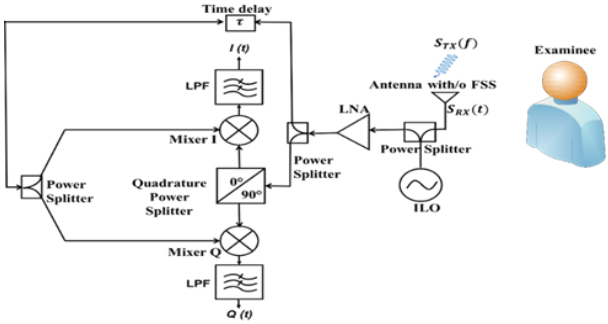

Fig. 4. Block diagram of the SIL radar system.

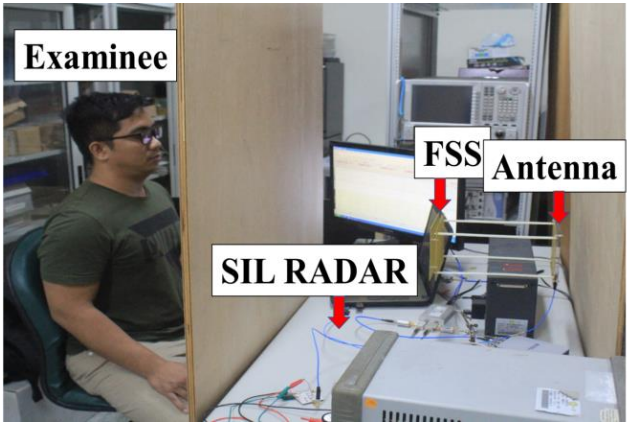

Fig. 5. Experimental setup of the vital sign monitoring through the wall.

Fig. 5 shows the experimental setup for monitoring vital sign at the operating frequency of 2.4 GHz. The subject was asked to sit in a stationary position in front of the antenna and there is a wall between the subject and the radar. Notably, the subject can't make a huge motion which will interference the vital sign detection.

The first signal is radiated by the self-injection-locked oscillator (SILO) and due to the SIL mechanism, the signal is then frequency modulated (FM) by the movement of the subject chest caused by the respiration and heartbeat. Mathematically, the instantaneous frequency modulation of the SILO can be calculated as

$$\Delta\omega_{SILO}(t) = -\omega_{LR}\sin\left(\frac{4\pi(R+x(t))}{\lambda}\right) \quad (1)$$

where $\omega_{LR}$ is the locking range of the SILO; $R$ specifies the distance between the antenna and the chest; $x(t)$ represents the instantaneous displacement of the chest, and $\lambda$ denotes the wavelength of the carrier wave. The vital signs can be extracted from the phase of the baseband IQ signal that shown in Fig. 4.

IV. EXPERIMENTAL RESULTS OF VITAL SIGN MONITORING

In the experiments, the subject was asked to sit in front of the antennas at a distance of 75 cm. Fig. 6 (a) and (b) shows the experimental result of the vital sign monitoring through the wall in a time domain and its spectrum, respectively. With the help of FSS, the antenna gain is improved thus the radar sensitivity also increased. The radar system can detect the subject's vital sign clearly. The subject has the respiration rate and heartbeat rate at 0.46 and 1.56 Hz, respectively.

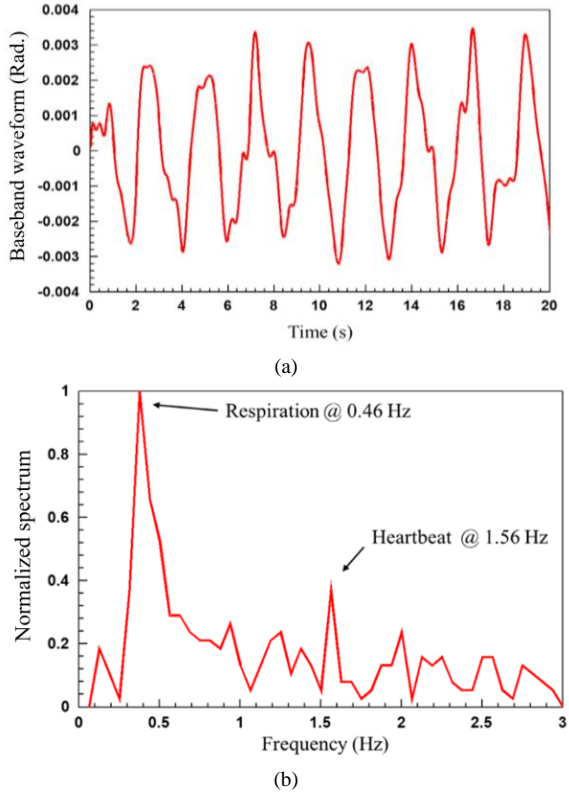

Fig. 6. Detected output spectrum of the wearable vital sign monitoring system. (a) Baseband waveform. (b) Normalized spectrum of Fig. 6(a).

## V. Conclusion

In this paper, the author successfully designs a 2.4 GHz phased array antenna 2×2 with frequency selective surface (FSS) to obtain a higher gain and efficiency. The FSS is designed with double-layer with unit cells of 9×7 for each layer. To enhance the gain, the author used a modified Jerusalem shaped cross then applied it to the upper layer of the antenna array. The total dimension of the antenna is 20×20×27.9 cm$^3$. Furthermore, the measurement of an antenna array efficiency is 65% with simulated 80% at the frequency of 2.4 GHz and 0.14 $\lambda_0$×0.14 $\lambda_0$ is the electrical dimensions of the unit cells. Vital sign monitoring of the subject has been done with the wall placed between the radar system amd the subject itself. The SIL radar can sucsessfully detect the subject respiration and heart beat clearly.